\documentclass[11pt]{article}
\pdfoutput=1
\usepackage[whole]{bxcjkjatype}
\usepackage[utf8]{inputenc}
\usepackage[
 a4paper,
 total={160mm,257mm},
 left=25mm,
 top=20mm]{geometry}
\usepackage{hyperref}
\usepackage{amsfonts}
\usepackage{amsmath}
\usepackage{amssymb}
\usepackage{mathrsfs}  
\usepackage{color}
\usepackage{authblk}
\usepackage{physics}
\usepackage{multicol}
\usepackage{tikz}

\newcommand{\ee}{\mathrm{e}}
\newcommand{\ii}{\mathrm{i}}

\makeatletter

\@addtoreset{equation}{section}
\makeatother

\title{\textbf{Universal edge scaling in random partitions}} 
\author{\textsc{Taro Kimura} and \textsc{Ali Zahabi}}
\affil{Institut de Math\'ematiques de Bourgogne,\\ Universit\'e Bourgogne Franche-Comt\'e, France}
\date{}

\begin{document}

\maketitle

\begin{abstract}
We establish the universal edge scaling limit of random partitions with the infinite-parameter distribution called the Schur measure.
We explore the asymptotic behavior of the wave function, which is a building block of the corresponding kernel, based on the Schr\"odinger-type differential equation.
We show that the wave function is in general asymptotic to the Airy function and its higher-order analogs in the edge scaling limit.
We construct the corresponding higher-order Airy kernel and the Tracy--Widom distribution from the wave function in the scaling limit, and discuss its implication to the multicritical phase transition in the large size matrix model.
We also discuss the limit shape of random partitions through the semi-classical analysis of the wave function. 
\end{abstract}

\tableofcontents

\section{Introduction}\label{sec:intro}

The paradigm of the universal distribution arising from the large size random matrix has been playing a central role at the crossroad of various research fields in both physics and mathematics~\cite{Mehta:2004RMT,Forrester:2010,Kuijlaars:2011}.
In particular, the statistical behavior of the largest eigenvalue is described by the Tracy--Widom distribution~\cite{Tracy:1992rf}, which is constructed by the Fredholm determinant of the universal edge scaling kernel, a.k.a., the Airy kernel~\cite{Bowick:1991ky,Forrester:1993vtx,Nagao:1993JPSJ}.
It is a universal distribution in the sense that one can obtain the Tracy--Widom distribution from totally different microscopic description by taking the proper scaling limit.
It is now known that the Tracy--Widom distribution is widely found in various contexts, including, the longest increasing subsequence of random permutations~\cite{Baik:1999JAMS}, the fluctuation in the (totally) asymmetric simple exclusion process ((T)ASEP)~\cite{Johansson:2000CMP}, the large $N$ phase transition of matrix model and gauge theory (Gross--Witten--Wadia/Douglas--Kazakov phase transition~\cite{Gross:1980he,Wadia:1980cp,Douglas:1993iia}. See also \cite{Majumdar:2013JSM, zahabi2016new}), integrable systems such as spin chains \cite{saeedian2018phase, saeedian2020exact}, and the growing turbulent interfaces~\cite{Praehofer:2000PRL,Takeuchi:2010PRL,Takeuchi:2011SR}.

The random partition, which is a main topic discussed in this paper, is interpreted as a discretized version of the random matrix~\cite{Kerov:2003,Olshanski:2011}.
From the Plancherel measure and its Poissonization, and also the infinite parameter generalization, a.k.a., the Schur measure~\cite{Okounkov:2001SM}, one can find a quite similar structure in the corresponding distribution and the correlation function to the random matrix.
In fact, the random partition has a realization as a special class of the unitary random matrix, which is interpreted as a consequence of the Schur--Weyl duality.
In this context, the matrix size $N$ is related to the constraint for the largest entry of the partition, and hence, the matrix integral is described as a special case of the gap probability based on a discretized version of the Fredholm determinant~\cite{Borodin:2000IEOT}.
Taking the scaling limit, this gap probability is asymptotic to the Tracy--Widom distribution, and in this sense, it belongs to the same universality class to the random matrix in the edge scaling limit.
However, such a connection between the random partition and the universal distribution has been so far established only for the specific case, i.e., the Plancherel measure with a single parameter, which is identified with the coupling constant of the unitary matrix model.

The purpose of this paper is to establish the universality of the edge scaling limit in more generic random partitions.
We show that such a universal behavior arises in the proper scaling limit of the Schur measure as a generalization of the Plancherel measure.
Furthermore, due to infinitely many parameters, one can consider the higher-order/multicritical scaling limit for the Schur measure random partition, and the higher-order version of the Tracy--Widom distribution is consequently obtained~\cite{Periwal:1990qb,Claeys:2009CPAM,LeDoussal:2018dls,Cafasso:2019IMRN}:
The explicit forms of the kernels and the asymptotics of the associated Fredholm determinants were obtained in~\cite{LeDoussal:2018dls} by considering the momentum distribution associated with the non-harmonic potential.
Then, the asymptotic behavior of the higher-order Tracy--Widom distribution was established in~\cite{Cafasso:2019IMRN} based on the Riemann--Hilbert analysis of the Painlev\'e II hierarchy.
We emphasize that our approach studied in this paper is to consider the scaling limit of the wave function, which is a building block of the corresponding kernel.
This provides a concise derivation of the higher-order Airy function and the corresponding Tracy--Widom distribution.

Such a higher-order behavior has been discussed in the gap closing regime of the random matrix, and we see that it naturally appears at the spectral edge of random partitions with the parameter tuning.
Since this universal behavior arises in the scaling limit of random partitions, the higher-order edge scaling behavior discussed in this paper is expected to be observed in wide-ranging class of statistical phenomena.

The remaining part of this paper is organized as follows.
We first introduce the Schur measure, which characterizes the distribution of random partitions, and explain that the corresponding correlation function is concisely described by the determinant of the kernel.
Then, we focus on the wave function, which is a building block of the kernel. 
We analyze the Schr\"odinger-type differential/difference equation, and show that the wave function is universally asymptotic to the Airy function and its higher-order analog in the edge scaling limit.
Based on this asymptotic behavior of the wave function, we derive the universal edge kernel, a higher-order version of the Airy kernel.
We then construct the higher-order Tracy--Widom distribution constructed with the higher-order Airy kernel, and discuss the associated multicritical behavior of the large size matrix model.
We demonstrate our formalism with the primary example of the random partition described by the Plancherel measure.
We also discuss a geometric aspect of the wave function, and show that the limit shape of the random partition is described by the semi-classical analysis of the Schr\"odinger-type equation.

\subsubsection*{Note added}

During finalizing the manuscript, we have noticed that the authors of the paper~\cite{Betea:2020} also discuss the multicritical behavior of random partitions through a different approach from ours.
We are grateful to D. Betea, J. Bouttier, and H. Walsh for their kind correspondence.

\section{Random partition with Schur measure}

\begin{figure}[t]
    \begin{center}
     \begin{tikzpicture}[scale=.5]
    \draw (-9,9) -- (0,0) -- (9,9);
    \foreach \x in {0,1,...,6}{
    \draw (\x,\x) -- ++(1,1) -- ++(-1,1) -- ++(-1,-1) -- cycle;        
    }
    \foreach \x in {0,1,...,5}{
    \draw (\x-1,\x+1) -- ++(1,1) -- ++(-1,1) -- ++(-1,-1) -- cycle;        
    }
    \foreach \x in {0,1,...,3}{
    \draw (\x-2,\x+2) -- ++(1,1) -- ++(-1,1) -- ++(-1,-1) -- cycle;        
    }
    \foreach \x in {0,1}{
    \draw (\x-3,\x+3) -- ++(1,1) -- ++(-1,1) -- ++(-1,-1) -- cycle;        
    }            
    \draw (-4,4) -- ++(1,1) -- ++(-1,1) -- ++(-1,-1) -- cycle;
    \draw (-5,5) -- ++(1,1) -- ++(-1,1) -- ++(-1,-1) -- cycle;
    \begin{scope}[shift={(0,-.5)}]
     \node at (9.5,0) {$\cdots$};
     \draw (8.5,0) circle (.3);    
     \draw (7.5,0) circle (.3);    
     \filldraw (6.5,0) circle (.3);
     \draw (5.5,0) circle (.3);
     \filldraw (4.5,0) circle (.3);
     \draw (3.5,0) circle (.3);
     \draw (2.5,0) circle (.3);
     \filldraw (1.5,0) circle (.3);
     \draw (.5,0) circle (.3);
     \draw (-.5,0) circle (.3);
     \filldraw (-1.5,0) circle (.3);
     \draw (-2.5,0) circle (.3);
     \filldraw (-3.5,0) circle (.3);
     \filldraw (-4.5,0) circle (.3);
     \draw (-5.5,0) circle (.3);
     \filldraw (-6.5,0) circle (.3);
     \filldraw (-7.5,0) circle (.3);
     \filldraw (-8.5,0) circle (.3);
     \node at (-9.5,0) {$\cdots$};     
    \end{scope}
    \draw [dotted] (6.5,0) -- ++(0,7.5);
    \draw [dotted] (4.5,0) -- ++(0,7.5);
    \draw [dotted] (1.5,0) -- ++(0,6.5);
    \draw [dotted] (-1.5,0) -- ++(0,5.5);
    \draw [dotted] (-3.5,0) -- ++(0,5.5);
    \draw [dotted] (-4.5,0) -- ++(0,6.5);
    \draw [dotted] (-6.5,0) -- ++(0,6.5);
    \draw [dotted] (-7.5,0) -- ++(0,7.5);
    \draw [dotted] (-8.5,0) -- ++(0,8.5);                        
     \end{tikzpicture}
    \end{center}
   \caption{The map from the partition $\lambda = (7,6,4,2,1,1)$ to the sequence of Maya particles at $\displaystyle {X}(\lambda) = \qty(\frac{13}{2},\frac{9}{2},\frac{3}{2},-\frac{3}{2},-\frac{7}{2},-\frac{9}{2})$.}
   \label{eq:Young_764211}
   \end{figure}
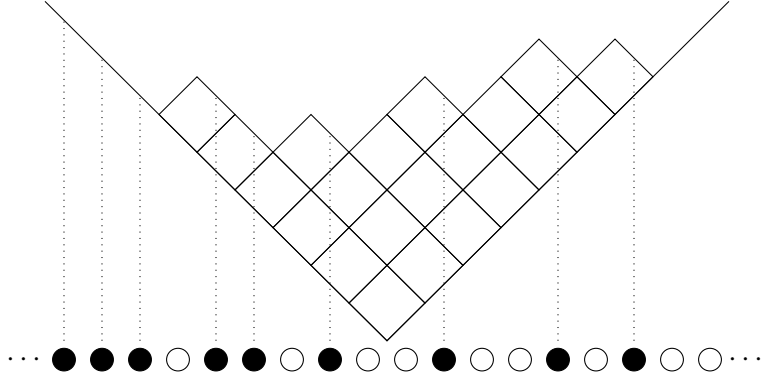

Let $\mathscr{Y}$ be a set of the partitions.
We consider the random distribution of partitions with the following measure, called the Schur measure~\cite{Okounkov:2001SM}:
\begin{align}
    \mu(\lambda) = \frac{1}{Z} \, s_\lambda(\mathsf{X}) s_\lambda(\mathsf{Y})
\end{align}
where $s_\lambda$ is the Schur function, and $\mathsf{X} = (\mathsf{x}_i)_{i \in \mathbb{N}}$ and $\mathsf{Y} = (\mathsf{y}_i)_{i \in \mathbb{N}}$ are sets of the parameters.
We also apply another parametrization using the Miwa variables, 
\begin{align}
    t_n = \frac{1}{n} \sum_{i=1}^\infty \mathsf{x}_i^n
    \, , \qquad
    \tilde{t}_n = \frac{1}{n} \sum_{i=1}^\infty \mathsf{y}_i^n
    \, .
    \label{eq:Miwa_var}
\end{align}
The constant $Z$ is the partition function obtained via the Cauchy sum formula
\begin{align}
    Z 
    = \sum_{\lambda \in \mathscr{Y}} s_\lambda(\mathsf{X}) s_\lambda(\mathsf{Y}) 
    = \prod_{1 \le i, j \le \infty} \qty( 1 - \mathsf{x}_i \mathsf{y}_j )^{-1} 
    = \exp \qty( \sum_{n = 1}^\infty n \, t_n \tilde{t}_n )
    \, ,
    \label{eq:Schur_part_fn}
\end{align}
so that the measure is normalized, $\sum_{\lambda \in \mathscr{Y}} \mu(\lambda) = 1$.
We then define the expectation value of the observable $\mathcal{O}(\lambda)$ with this measure as
\begin{align}
    \expval{ \mathcal{O}(\lambda) } = \sum_{\lambda \in \mathscr{Y}} \mu(\lambda) \, \mathcal{O}(\lambda)
    \, .
\end{align}
Then, the $k$-point correlation function is defined with a set $W = (w_i)_{i = 1,\ldots,k} \subset \mathbb{Z} + \frac{1}{2}$ as follows:
\begin{align}
    \rho_k(W) 
    = \expval{ \prod_{w \in W} \delta_{w}({X}(\lambda)) }
  \label{eq:rho_fn}
\end{align}
with the ``density function'' 
\begin{align}
    \delta_w (\mathscr{X}) =
    \begin{cases}
    1 & (w \in \mathscr{X}) \\ 0 & (w \not\in \mathscr{X})
    \end{cases}
\end{align}
and the boson-fermion map,%
\footnote{%
Not to be confused with the Schur measure parameter $\mathsf{X} = (\mathsf{x}_i)_{i \in \mathbb{N}}$.
}
\begin{align}
    {X}(\lambda) = 
    \qty( x_i = \lambda_i - i + \frac{1}{2} )_{i \in \mathbb{N}}
    \subset \mathbb{Z} + \frac{1}{2}
    \, .
    \label{eq:Maya}
\end{align}
We remark that $(x_i)_{i \in \mathbb{N}}$ are interpreted as the coordinates of the Maya particles 
\tikz[baseline={([yshift=-8pt]current bounding box.north)},thick] \filldraw [fill=black] (0,0) circle (.15); 
as shown in Fig.~\ref{eq:Young_764211}.
See~\cite{Jimbo:1983if} for details.

It has been shown that the correlation function~\eqref{eq:rho_fn} is described as a size $k$ determinant~\cite{Okounkov:2001SM}
\begin{align}
    \rho_k(W) = \det_{1 \le i, j \le k} K(w_i, w_j)
    \, ,
\end{align}
where the kernel is given by 
\begin{subequations}\label{eq:Schur_kernel}
\begin{align}
    K(r,s) & = 
    \frac{1}{(2 \pi \ii)^2}
    \oint_{|z|>|w|} \hspace{-1.5em} \dd{z} \dd{w} \,
    \frac{\mathsf{K}(z,w)}{z^{r + 1/2} w^{- s + 1/2}}
    \qquad \qty(r, s \in \mathbb{Z} + \frac{1}{2})
    \, , \\
    \mathsf{K}(z,w) & = \frac{\mathsf{J}(z)}{\mathsf{J}(w)} \, \frac{1}{z - w}
    \, , \qquad
    \mathsf{J}(z) = \prod_{n = 1}^\infty \frac{1 - \mathsf{x}_n z}{1 - \mathsf{y}_n / z }
    \, .
\end{align}
\end{subequations}
We call $\mathsf{J}(z)$ the wave function in the following.

\section{Wave function analysis}

Together with the Miwa variable~\eqref{eq:Miwa_var}, we apply the mode expansion to the wave function,
\begin{align}
    \mathsf{J}(z) 
    = \exp \left[ \sum_{n = 1}^\infty \qty( t_n \, z^n - \tilde{t}_n \, z^{-n}) \right]
    = \sum_{x \in \mathbb{Z}} J(x) \, z^{x}
    \, , \qquad
    J(x) =  \oint \frac{\dd{z}}{2 \pi \ii} \frac{\mathsf{J}(z)}{z^{x+1}}
    \, .
    \label{eq:J_exp}
\end{align}
We call the mode coefficients $(J(x))_{x \in \mathbb{Z}}$ the wave functions as well, which are interpreted as a multi-variable generalization of the Bessel function~\cite{Okounkov:2001SM}.

One can show that the wave functions satisfy the following differential and difference equations:
\begin{subequations}\label{eq:ODE_J}
\begin{align}
    \qty[
    \sum_{n = 1}^\infty n \qty( t_n z^n + \tilde{t}_n z^{-n}) - z \pdv{}{z}
    ] \mathsf{J}(z) & = 0
    \label{eq:ODE1}
    \\[.5em]
    \quad \iff \quad
    \qty[
    \sum_{n = 1}^\infty n  \qty( t_n \nabla_x^n + \tilde{t}_n \nabla_x^{-n} ) - x
    ] J(x) & = 0
    \label{eq:ODE2}
\end{align}
\end{subequations}
where we define the shift operator $\nabla_x f(x) = f(x+1)$ with $\nabla_x = \exp \qty( \partial_x )$.
In fact, the $(x,z)$ variables are converted to each other through the Fourier transformation, $(z,\partial_{\log z}) \leftrightarrow (\ee^{\partial_x}, x)$, as seen from the relation~\eqref{eq:J_exp}.

Let us discuss the scaling limit of the difference equation for the wave function~\eqref{eq:ODE2}.
We introduce the parameter $\epsilon$, which rescales the variables $(x,t_n, \tilde{t}_n) \to (x/\epsilon,t_n/\epsilon,\tilde{t}_n/\epsilon)$.
Then, expanding the shift operator,
\begin{align}
    \nabla_{x/\epsilon}^n = \sum_{k = 0}^\infty \frac{(n \epsilon)^k}{k!} \dv[k]{}{x}
    \, , \qquad
    \nabla_{x/\epsilon}^{-n} = \sum_{k = 0}^\infty \frac{(- n \epsilon)^k}{k!} \dv[k]{}{x}    
    \, ,
\end{align}
we obtain an alternative form of the difference equation in terms of infinitely many differentials,
\begin{align}
    \qty[
    \sum_{k = 1}^\infty \alpha_{k} \, \epsilon^{k} \dv[k]{}{x} - (x - \beta)
    ] J\qty(\frac{x}{\epsilon}) = 0
    \, ,
\end{align}
where we define the new coefficients
\begin{align}
    \alpha_k = \sum_{n=1}^\infty \frac{n^{k+1}}{k!} \, \qty( t_n + (-1)^k \tilde{t}_n )
    \, , \qquad
    \beta = \alpha_0 = \sum_{n=1}^\infty n \, (t_n + \tilde{t}_n)
    \, .
    \label{eq:alpha_beta}
\end{align}

In order to discuss the scaling limit of the difference equation, we introduce the scaling variable in the vicinity of the spectral edge $x \approx \beta$ (see the discussion around~\eqref{eq:sp_edge}),
\begin{align}
    x = \beta + \alpha_p^{\frac{1}{p+1}} \epsilon^{\frac{p}{p+1}} \xi
    \qquad \text{for} \qquad
    p \ge 2
    \, .
    \label{eq:sc_var}
\end{align}
Taking the limit $\epsilon \to 0$ with keeping $\alpha_{p'} = 0$ for $p' < p$, we obtain the differential equation,
\begin{align}
    \qty[ \dv[p]{}{\xi} - \xi ] J\qty(\frac{\beta}{\epsilon} + \qty( \frac{\alpha_{p}}{\epsilon})^{\frac{1}{p+1}} \xi ) = 0
    \, .
\end{align}
In general, one can also incorporate the lower derivative terms, $(\partial_\xi)^{p'}$ for $p'<p$, by tuning the parameters.
Therefore, the scaling limit of the wave function is given by the $p$-Airy function,
\begin{align}
    J\qty(\frac{\beta}{\epsilon} + \qty( \frac{\alpha_{p}}{\epsilon})^{\frac{1}{p+1}} \xi )
    \ \xrightarrow{\epsilon \to 0} \
    \operatorname{Ai}_p(\xi) = \int_\gamma \dd{x} \exp\qty( (-1)^p \frac{x^{p+1}}{p+1} - x \xi )
    \, ,
\end{align}
where $\gamma$ is an integral contour providing a converging integral.
The case with $p = 2$ corresponds to the standard Airy function, $\operatorname{Ai}_2(x) = \operatorname{Ai}(x)$.
We emphasize that this asymptotic behavior is universal in the sense that it does not depend on the parameter $(t_n,\tilde{t}_n)_{n \in \mathbb{N}}$ characterizing the microscopic distribution of random partitions.

\section{Kernel analysis}

Hereafter, we assume $\mathsf{x}_i = \mathsf{y}_i$ for $\forall i \in \mathbb{N}$ (equivalently, $t_n = \tilde{t}_n$ for $\forall n \in \mathbb{N}$).
Then, the wave function exhibits the property $\mathsf{J}(z^{-1}) = \mathsf{J}(z)^{-1}$.
In this case, applying the formula~\eqref{eq:Schur_kernel}, we arrive at the expression in terms of the wave functions,
\begin{subequations}
\begin{align}
    & \mathsf{K}(z,w) = \sum_{n,m \in \mathbb{Z}} \sum_{k = 1}^\infty J(n) J(m) \, z^{n - k } w^{- m + k - 1}
    \\
    \ \iff \
    & K(r,s) = \sum_{k = 1}^\infty J\qty(r + k - \frac{1}{2}) J\qty(s + k - \frac{1}{2})
    \, .
\end{align}
\end{subequations}
We can show that the kernel is projective
\begin{align}
    \sum_{u \in \mathbb{Z} + 1/2} K(r, u) K(u, s) = K(r,s)
    \, ,
\end{align}
using the orthonormal condition of the wave function for the case $t_n = \tilde{t}_n$,
\begin{align}
    \sum_{k \in \mathbb{Z}} J(n+k) J(m+k) = \delta_{n,m}
    \, .
    \label{eq:orthonormal_J}
\end{align}

Let us then discuss the scaling limit of the kernel.
Since we impose the condition $t_n = \tilde{t}_n$, the odd coefficients become zero, $\alpha_{k} = 0$ for $k \in 2 \mathbb{Z}_{\ge 0} + 1$, as seen from their definition~\eqref{eq:alpha_beta}.
Therefore, we take the scaling limit with the scaling variable \eqref{eq:sc_var} for $p \in 2 \mathbb{N}$:
\begin{align}
    & 
    K\qty( \frac{\beta}{\epsilon} + \qty( \frac{\alpha_{p}}{\epsilon})^{\frac{1}{p+1}} x,\, \frac{\beta}{\epsilon} + \qty( \frac{\alpha_{p}}{\epsilon})^{\frac{1}{p+1}} y)
    \nonumber \\ &
    = \sum_{k = 1}^\infty J\qty(\frac{\beta}{\epsilon} + \qty( \frac{\alpha_{p}}{\epsilon})^{\frac{1}{p+1}} x + k - \frac{1}{2}) J\qty(\frac{\beta}{\epsilon} + \qty( \frac{\alpha_{p}}{\epsilon})^{\frac{1}{p+1}} y + k - \frac{1}{2})
    \nonumber \\
    & \xrightarrow{\epsilon \to 0} \
    \qty( \frac{\alpha_{p}}{\epsilon})^{\frac{1}{p+1}}
    K_{p\text{-Airy}}(x,y)
\end{align}
where the $p$-Airy kernel is defined~\cite{Brezin:1998zz,Brezin:1998PREb,LeDoussal:2018dls}
\begin{subequations}
\begin{align}
    K_{p\text{-Airy}}(x,y) 
    & = \int_0^\infty \dd{z} \operatorname{Ai}_p(x+z) \operatorname{Ai}_p(y+z)
    \\
    & = \frac{1}{x - y} \sum_{q=0}^{p-1} (-1)^{q} \operatorname{Ai}_p^{(q)}(x) \operatorname{Ai}_p^{(p-q-1)}(y)
    \, .
    \label{eq:Airy_kernel2}
\end{align}
\end{subequations}
We denote the $r$-th derivative of the Airy function by $\operatorname{Ai}_p^{(r)}(x)$.
In order to obtain the expression~\eqref{eq:Airy_kernel2}, it will be helpful to use the relation
\begin{align}
    \dv{}{z} K_{p\text{-Airy}}(x+z,y+z) 
    = - \operatorname{Ai}_p(x+z) \operatorname{Ai}_p(y+z)
    \, .
\end{align}
One can also show that the $p$-Airy kernel is projective
\begin{align}
    \int_{- \infty}^\infty \dd{z} K_{p\text{-Airy}}(x,z) K_{p\text{-Airy}}(z,y) = K_{p\text{-Airy}}(x,y)
    \, ,
\end{align}
using the orthonormality of the $p$-Airy function,
\begin{align}
    \int_{- \infty}^\infty \dd z \operatorname{Ai}_p(x + z) \operatorname{Ai}_p(y + z) = \delta(x - y)
    \, .
\end{align}
For example, the lower degree cases are given as~\cite{LeDoussal:2018dls},
\begin{subequations}
\begin{align}
    K_{2\text{-Airy}}(x,y) & = \frac{\operatorname{Ai}(x) \operatorname{Ai}'(y) - \operatorname{Ai}'(x) \operatorname{Ai}(y)}{x - y}
    \, , \label{eq:A_kernel2} \\
    K_{4\text{-Airy}}(x,y) & = \frac{\operatorname{Ai}_4(x) \operatorname{Ai}_4'''(y) - \operatorname{Ai}_4'(x) \operatorname{Ai}_4''(y) + \operatorname{Ai}_4''(x) \operatorname{Ai}_4''(y) - \operatorname{Ai}_4'''(x) \operatorname{Ai}_4'(y)}{x - y}
    \, .
\end{align}
\end{subequations}

\section{Higher-order Tracy--Widom distribution}

Once the kernel for the correlation function is obtained, one can systematically consider the gap probability:
For the integral operator defined with the kernel on a specific domain $I$, $\displaystyle (\hat{K} \cdot f)_I(x) = \int_I \dd{y} K(x,y) f(y)$, the probability such that any particles are not found in $I$ is given by the Fredholm determinant of the kernel,
\begin{align}
    \det( 1 - \hat{K})_I = \sum_{n = 0}^\infty \frac{(-1)^n}{n!} \int_{I^{n}} \prod_{i=1}^n \dd{x_i} \det_{1 \le i, j \le n} K(x_i, x_j)
    \, .
\end{align}
In the context of the random matrix theory, we are in particular interested in the gap probability with the Airy kernel with $I = (s, \infty)$, a.k.a., the Tracy--Widom distribution, which describes the statistical behavior of the largest eigenvalue in the scaling limit~\cite{Tracy:1992rf},
\begin{align}
    F(s) = \det( 1 - \hat{K}_\text{Airy})_{(s,\infty)}
    \, .
\end{align}
A similar description is also possible for random partitions, which is a discrete analog of the random matrix, and in this context, the largest eigenvalue is replaced with the largest Maya particle, which corresponds to the first entry of the partition $x_1 = \lambda_1 - 1/2$ as in \eqref{eq:Maya} (the right most particle
\tikz[baseline={([yshift=-8pt]current bounding box.north)},thick] \filldraw [fill=black] (0,0) circle (.15);
in Fig.~\ref{eq:Young_764211})~\cite{Borodin:2000IEOT}.
Then, applying the same argument together with the scaling variable~\eqref{eq:sc_var}, we arrive at the following statement for the higher-order edge scaling behavior:
\begin{align}
    \lim_{\epsilon \to 0} \mathbb{P}\qty[ \frac{\lambda_1 - \beta / \epsilon}{(\alpha_p / \epsilon)^{\frac{1}{p+1}}} < s] 
    = \det( 1 - \hat{K}_{p\text{-Airy}})_{(s,\infty)}
    =: F_p(s)
    \, ,
    \label{higher TW }
\end{align}
where $F_p(s)$ is a higher-order analog of the Tracy--Widom distribution~\cite{Periwal:1990qb,Claeys:2009CPAM,LeDoussal:2018dls,Cafasso:2019IMRN}.
We remark that in the scaling limit $\epsilon \to 0$, we should also tune the parameter such that $\alpha_{p'} = 0$ for $p' < p$.
See also the alternative derivation~\cite{Betea:2020}.

\subsection{Multicritical phase transition}

We briefly discuss a possible phase transition associated with the higher-order Tracy--Widom distribution.
Imposing the constraint on the largest entry of the partition, the summation over the random partition is rewritten as the unitary matrix integral~\cite{Borodin:2000IEOT}: 
\if0\footnote{%
Precisely speaking, we should impose the constraint $\lambda_1^\text{T} \le N$ to obtain the $\mathrm{U}(N)$ integral, which requires for redefinition of the Miwa variables $(t_n,\tilde{t}_n)_{n \in \mathbb{N}}$.
}\fi
\begin{align}
  \mathcal{Z}_N 
  = \sum_{\lambda_1 \le N} s_\lambda(\mathsf{X}) s_\lambda(\mathsf{Y})
  = \int_{\mathrm{U}(N)} \hspace{-1em} \dd{U} \exp\qty( \sum_{n=1}^\infty \qty( t'_n \tr U^n + \tilde{t}'_n \tr U^{-n}) ),
  \label{pf}
\end{align}
where we use another set of the Miwa variables compared to the previous case~\eqref{eq:Miwa_var},
\begin{align}
    t'_n = \frac{(-1)^{n-1}}{n} \sum_{i=1}^\infty \mathsf{x}_i^n
    \, , \qquad
    \tilde{t}'_n = \frac{(-1)^{n-1}}{n} \sum_{i=1}^\infty \mathsf{y}_i^n
    \, .
    \label{eq:Miwa_var_mod}
\end{align}

We define the free energy from the matrix integral,
\begin{align}
    \mathcal{F}=\lim_{\epsilon \to 0} \epsilon^2 \log  \mathcal{Z}_N/Z,
\end{align}
where $Z$ is the partition function of the Schur measure~\eqref{eq:Schur_part_fn} and  we fix $\gamma := N \epsilon = {O}(1)$. 
This is the 't Hooft limit of the $\mathrm{U}(N)$ matrix model.
Since, in the case with $t'_n \neq \tilde{t}'_n$, we do not have the expression for the kernel, we assume $t'_n = \tilde{t}'_n$ for the moment.
Then, using Eq.~\eqref{higher TW } together with the wave function analysis~\eqref{eq:sc_var}, we observe that the free energy can be written for $p \in 2 \mathbb{N}$ as~\cite{LeDoussal:2018dls},
\begin{align}
    \mathcal{F}_p(s)=\lim_{\epsilon \to 0} \epsilon^2 \log F_p(s)
    \qquad \text{where} \qquad
    s= \frac{ (\gamma-\beta)/\epsilon}{(\alpha_p/\epsilon)^{\frac{1}{p+1}}}.
\end{align}
Although we have only discussed the even order Tracy--Widom distribution ($p \in 2 \mathbb{N}$), it suggests that the expression of the scaling variable $s$ is available for generic $p$.

We claim that, in the system defined by the partition function \eqref{pf}, there is a multicritical phase transition 
at the critical point $\beta_c=\gamma$, which turns out to be the spectral edge~\eqref{eq:sp_edge}. These phase transitions stem from the leading order asymptotic behavior of the higher-order Tracy--Widom distribution, which is conjectured to behave, for generic $p$ and up to some numerical factors, as
\begin{align}
     \lim_{s\to -\infty} \log F_p(s) \sim |s|^{\frac{2(p+1)}{p}}, \quad  \lim_{s\to \infty} \log F_p(s) \sim
    \log \left( 1 - s^{-\frac{p+1}{p}} \exp({-s^{\frac{p+1}{p}}}) \right).
\end{align}
The above conjecture is consistent with the asymptotic behavior of the Tracy--Widom distribution ($p=2$)~\cite{baik2008asymptotics}, and recent asymptotic results for the Pearcey processes
($p=3$)~\cite{dai2020asymptotics}.
See also earlier results~\cite{Brezin:1998zz,Brezin:1998PREb}.

Using the above leading asymptotic behavior of the higher-order Tracy--Widom distribution, we observe that the leading order, non-zero contribution to the free energy in the regime $s\to -\infty$ ($\gamma<\beta$) is given by
\begin{align}
    \mathcal{F}_p(\gamma)= \alpha_p^{-2/p}(\gamma-\beta)^{2(p+1)/p}+ O(\epsilon^2).
\end{align}
Comparing the above free energy in the phase $\gamma<\beta$, with the leading contribution to the free energy in the phase $\gamma>\beta$, which is zero, we observe that $(2(p+1)/p)$-th derivative of the free energy is discontinuous at $\beta_c=\gamma$ and thus there is a $(2(p+1)/p)$-th order multicritical phase transition, at this critical point. Notice that at $p=2$, this is a third-order phase transition of the Gross--Witten--Wadia model~\cite{Gross:1980he,Wadia:1980cp}, see also \cite{zahabi2016new}, and as we increase $p$, we encounter a fractional-order multicritical phase transition for $p>2$. It seems that the model is flowing from a fixed-point with phase structure of order three (at $p=2$) to another fixed point with phase structure of order two at infinite $p$ and as we flow from the former fixed-point to the later fixed point, the fractional-order is monotonically decreasing from three to two. 

\section{Example: Plancherel measure}

Let us study the simplest example with a single parameter $t_1 = \tilde{t}_1 = \mathfrak{q}$ with $t_n = \tilde{t}_n = 0$ for $n \ge 2$.
In this case, the Schur measure is reduced to the Poissonized Plancherel measure,
\begin{align}
    \mu(\lambda) \ \xrightarrow{t_n = \tilde{t}_n = \mathfrak{q} \, \delta_{n,1}} \
    \mu_\text{PP}(\lambda) = \frac{1}{Z} \, \mathfrak{q}^{2|\lambda|} \qty( \frac{\dim \lambda}{|\lambda|} )^2,
\end{align}
where $|\lambda| = \sum_{i = 1}^\infty \lambda_i$, and $\dim \lambda$ is the dimension of the irreducible representation parametrized by $\lambda$ of the symmetric group $\mathfrak{S}_\infty$.
The partition function~\eqref{eq:Schur_part_fn} is given by $\displaystyle Z = \exp \qty(\sum_{n=1}^\infty n t_n \tilde{t}_n) = \ee^{\mathfrak{q}^2}$ in this case. 

The wave function is then given by the Bessel function $J(x) = J_x(2 \mathfrak{q})$ with the generating function
\begin{align}
    \mathsf{J}(z) 
    = \ee^{\mathfrak{q} (z - z^{-1})}
    = \sum_{x \in \mathbb{Z}} J_x (2 \mathfrak{q}) \, z^x
    \, .
\end{align}
Since the Bessel function obeys the difference equation, as a special case of \eqref{eq:ODE2},
\begin{align}
    \qty[ \nabla_x + \nabla_x^{-1} - \frac{x}{\mathfrak{q}} ] J_x(2\mathfrak{q}) = 0
    \, ,
    \label{eq:ODE_B}
\end{align}
we obtain the discrete Bessel kernel~\cite{Brodin:2000} 
\begin{align}
    K(r,s) = \mathfrak{q} \frac{J_{r - 1/2}(2\mathfrak{q}) J_{s + 1/2}(2\mathfrak{q}) - J_{r+1/2}(2\mathfrak{q}) J_{s - 1/2}(2\mathfrak{q})}{r-s}
    \, .
\end{align}
Let us discuss the scaling limit of the wave function and the kernel in the following.

\subsection{Bulk scaling limit}

Let us discuss the scaling limit of the Bessel function.
We first consider a bit different scaling than before, $(x,\mathfrak{q}) \to (x,\mathfrak{q}/\epsilon)$, in which the difference equation~\eqref{eq:ODE_B} becomes
\begin{align}
    \qty[ \nabla_x + \nabla_x^{-1} + O(\epsilon) ] J_n\qty(2\mathfrak{q}/\epsilon) = 0    
    \, .
\end{align}
Hence, the Bessel function behaves as the plane wave in the limit $\epsilon \to 0$:
\begin{align}
    J_x(2\mathfrak{q}/\epsilon) \ \xrightarrow{\epsilon \to 0} \ 
    \begin{cases}
    \displaystyle
    \cos\qty(\frac{\pi}{2} x) = (-1)^{x/2} & (x \in 2 \mathbb{Z}) \\[1em]
    \displaystyle
    \sin\qty(\frac{\pi}{2} x) = (-1)^{x/2-1/2} & (x \in 2 \mathbb{Z} + 1) 
    \end{cases}
\end{align}
We remark the relation $J_{-x}(z) = (-1)^x J_x(z)$.
Then, the discrete Bessel kernel is asymptotic to the sine kernel
\begin{align}
    K(r,s) \ \xrightarrow{\epsilon \to 0} \ \frac{\sin \pi (r - s)/2}{\pi(r - s)/2}
    \, ,
\end{align}
where the normalization is fixed to be $K(r,r) = 1$.

\subsection{Edge scaling limit}

We then consider the edge scaling limit.
The difference equation~\eqref{eq:ODE_B} is described with the scaled parameters $(x,\mathfrak{q}) \to (x/\epsilon,\mathfrak{q}/\epsilon)$ as follows:
\begin{align}
 \qty[ \epsilon^2 \dv[2]{}{x} - \qty( \frac{x}{\mathfrak{q}} - 2 ) + O(\epsilon^4) ] J_{x/\epsilon}(2\mathfrak{q}/\epsilon) = 0
 \, .
\end{align}
Therefore, the Bessel function is asymptotic to the Airy function,
\begin{align}
    J_{x/\epsilon}(2\mathfrak{q}/\epsilon)
    \ \xrightarrow{\epsilon \to 0} \
    \operatorname{Ai}(\xi)
    \qquad \text{with} \qquad
    x = 2 \mathfrak{q} + \epsilon^{2/3} \mathfrak{q}^{1/3} \xi
    \, ,
\end{align}
and the discrete Bessel kernel is given by the Airy kernel~\eqref{eq:A_kernel2} in the scaling limit,
\begin{align}
    \lim_{\epsilon \to 0}
    \qty(\frac{\epsilon}{\mathfrak{q}})^{1/3}
    K\qty(\frac{2\mathfrak{q}}{\epsilon} + \qty(\frac{\mathfrak{q}}{\epsilon})^{1/3} x, \frac{2\mathfrak{q}}{\epsilon} + \qty(\frac{\mathfrak{q}}{\epsilon})^{1/3} y)
    =
    K_\text{Airy}(x,y)
    \, .
\end{align}
We remark that, in this case, since there is only a single parameter $t_1 = \tilde{t}_1 = \mathfrak{q}$, one cannot realize the higher-order scaling limit $(p > 2)$.

\section{Spectral curve, semi-classical analysis, and limit shape}\label{sec:sp_curve}

The differential/difference equation for the wave function~\eqref{eq:ODE_J} is also useful to discuss the limit shape of the random partition~\cite{Logan1977206,Vershik:1977}.
Let us apply the scaled variables $(x,t_n,\tilde{t}_n) \to (x/\epsilon,t_n/\epsilon,\tilde{t}_n/\epsilon)$.
Then, the difference equation~\eqref{eq:ODE2} is written in the following form:
\begin{align}
    H(\hat{x},\hat{y}) J(x) = 0
    \, ,
    \label{eq:q_curve}
\end{align}
where the two-variable function is given by
\begin{align}
    H(x,y) = \sum_{n = 1}^\infty n \qty( t_n y^n + \tilde{t}_n y^{-n}) - x
    \, ,
    \label{eq:H_fn}
\end{align}
with the operator pair $(\hat{x},\hat{y})$ 
\begin{align}
    \hat{x} = x
    \, , \qquad
    \hat{y} = \nabla_{x/\epsilon} = \ee^{\epsilon \partial_x}
    \, .
    \label{eq:canonical_pair}
\end{align}
They are interpreted as the canonical operator pair obeying the canonical commutation relation
\begin{align}
    \comm{\log \hat{y}}{\hat{x}} = \epsilon
    \, .
    \label{eq:CCR}
\end{align}
It is also possible to describe the other differential equation~\eqref{eq:ODE1} by changing the operators from~\eqref{eq:canonical_pair} as $(\hat{x},\hat{y}) = (\partial_{\log z}, z)$ with the same canonical commutation relation~\eqref{eq:CCR} (up to rescaling with $\epsilon$).

In fact, the Schr\"odinger-type differential/difference equation written in the form of~\eqref{eq:q_curve} is interpreted as a quantization of the spectral curve, a.k.a., the quantum curve, defined as the zero locus of the algebraic function,
\begin{align}
    \Sigma = \qty{ (x,y) \in \mathbb{C} \times \mathbb{C}^\times \mid H(x,y) = 0 }
    \, ,
\end{align}
equipped with the differential one-form and the symplectic two-form
\begin{align}
    \lambda = \log y \, d x
    \, , \qquad
    \omega = d \lambda = d \log y \wedge d x
    \, .
\end{align}
We see that the canonical commutation relation~\eqref{eq:CCR} is taken with respect to the symplectic form $\omega$.
From this point of view, the scaling parameter $\epsilon$ plays a role of the Planck constant (quantum deformation parameter), and thus the scaling limit $\epsilon \to 0$ corresponds to the semi-classical limit of the quantum curve.
See~\cite{Eynard:2015aea} for details.

We can derive the limit shape of the random partition from the spectral curve as follows.
The variable $\log y$ is identified with the auxiliary function, called the resolvent, and its asymptotic form in the scaling limit $\epsilon \to 0$ is indeed determined by the algebraic equation, $H(x,y) = 0$.
Therefore, solving this zero locus equation, we obtain the limiting form of the resolvent.
For example, in the case of the Plancherel measure, $t_n = \tilde{t}_n = \mathfrak{q} \, \delta_{n,1}$, it is given as 
\begin{align}
    H(x,y) = y + y^{-1} - \frac{x}{\mathfrak{q}}
    \ \implies \
    \log y(x) = \operatorname{arccosh} \qty( \frac{x}{2 \mathfrak{q}} )
    \, .
\end{align}
In fact, since the degree of the variable $y$ is two, the spectral curve $\Sigma = \{ H(x,y) = 0 \}$ is a hyperelliptic curve in this case.
For generic case, it will not be the case since the degree of the $y$-variable becomes higher than two.

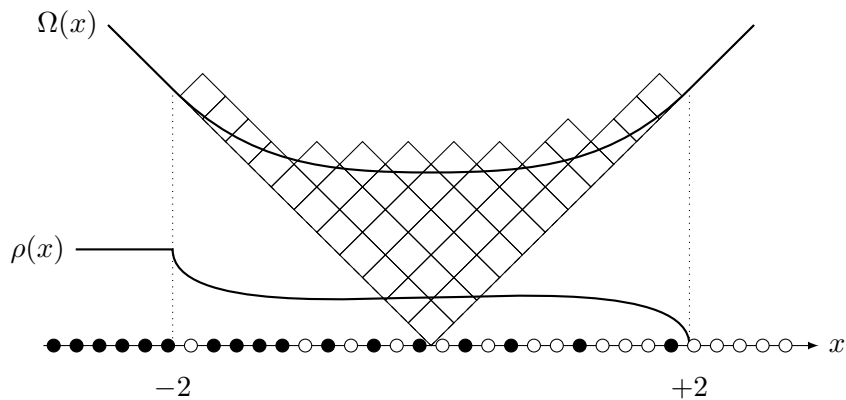
\begin{figure}[t]
\begin{center}
 \begin{tikzpicture}[scale=.85]

  \draw [densely dashed] (-5,5) node [left] {$\Omega(x)$} -- (0,0) -- (5,5);
  \draw [-latex] (-6,0) -- (6,0) node [right] {$x$};

  \draw [thick] (-5,5) -- (-4,4) to [out=-45,in=180] (0,2.7) to [out=0,in=225] (4,4) -- (5,5);

  \draw [dotted] (-4,4) -- (-4,0) node [below=.75em] {$-2$};
  \draw [dotted] (4,4) -- (4,0) node [below=.75em] {$+2$};  

  \draw [thick] (-5.5,1.5) node [left] {$\rho(x)$} -- (-4,1.5)
  .. controls ++(0,-1) and (-1,.75) .. (0,.75) .. controls ++(1,0) and (4,1) .. (4,0);

    \begin{scope}[rotate=45,scale=.5]
     
     \draw (0,0) rectangle ++(1,1);
     \draw (1,0) rectangle ++(1,1);
     \draw (2,0) rectangle ++(1,1);
     \draw (3,0) rectangle ++(1,1);
     \draw (4,0) rectangle ++(1,1);
     \draw (5,0) rectangle ++(1,1);
     \draw (6,0) rectangle ++(1,1);
     \draw (7,0) rectangle ++(1,1);
     \draw (8,0) rectangle ++(1,1);
     \draw (9,0) rectangle ++(1,1);
     \draw (10,0) rectangle ++(1,1);            
     
     \draw (0,1) rectangle ++(1,1);
     \draw (1,1) rectangle ++(1,1);
     \draw (2,1) rectangle ++(1,1);
     \draw (3,1) rectangle ++(1,1);
     \draw (4,1) rectangle ++(1,1);
     \draw (5,1) rectangle ++(1,1);
     \draw (6,1) rectangle ++(1,1);
     \draw (7,1) rectangle ++(1,1);   
     
     \draw (0,2) rectangle ++(1,1);
     \draw (1,2) rectangle ++(1,1);
     \draw (2,2) rectangle ++(1,1);
     \draw (3,2) rectangle ++(1,1);
     \draw (4,2) rectangle ++(1,1);
     \draw (5,2) rectangle ++(1,1);
     
     \draw (0,3) rectangle ++(1,1);
     \draw (1,3) rectangle ++(1,1);
     \draw (2,3) rectangle ++(1,1);
     \draw (3,3) rectangle ++(1,1);
     \draw (4,3) rectangle ++(1,1);
     
     \draw (0,4) rectangle ++(1,1);
     \draw (1,4) rectangle ++(1,1);
     \draw (2,4) rectangle ++(1,1);
     \draw (3,4) rectangle ++(1,1);
     
     \draw (0,5) rectangle ++(1,1);
     \draw (1,5) rectangle ++(1,1);
     \draw (2,5) rectangle ++(1,1);
     
     \draw (0,6) rectangle ++(1,1);
     \draw (1,6) rectangle ++(1,1);
     
     \draw (0,7) rectangle ++(1,1);
     \draw (0,8) rectangle ++(1,1);
     \draw (0,9) rectangle ++(1,1);
     \draw (0,10) rectangle ++(1,1);            
     
    \end{scope}


  \foreach \x in {-11,-6,-4,-2,0,2,4,5,7,8,9,11,12,13,14,15}{

  \filldraw[white,draw=black] (.354*\x+.177,0) circle (.1);    

  }

  \foreach \x in {-17,-16,-15,-14,-13,-12,-10,-9,-8,-7,-5,-3,-1,1,3,6,10}{

  \filldraw (.354*\x+.177,0) circle (.1);    

  }

 \end{tikzpicture}
\end{center}
   \caption{The Maya particle density function $\rho(x)$ and the limit shape $\Omega(x)$}
   \label{fig:limit_shape}
  \end{figure}

We reproduce the density function $\rho(x)$ from the resolvent for the Maya particles~\eqref{eq:Maya}.
Regarding the asymptotic form of the density function,
\begin{align}
    \rho(x) 
    \ \longrightarrow \
    \begin{cases}
    1 & (x \to - \infty) \\
    0 & (x \to + \infty) 
    \end{cases}
\end{align}
the density function is given by
\begin{align}
    \rho(x)
    =
    \begin{cases}
    1 & ( x < - 2 \mathfrak{q}) \\
    \displaystyle
    \frac{1}{\pi} \arccos\qty(x/2\mathfrak{q}) & ( |x| < 2 \mathfrak{q} ) \\
    0 & ( x > + 2 \mathfrak{q} )
    \end{cases}
\end{align}
After rescaling $x/\mathfrak{q} \to x$, the limit shape of the random partition $\Omega(x)$ is obtain from the density function (Fig.~\ref{fig:limit_shape}. See also~\cite{Kimura:2012iua}).
Since the Maya particle \tikz[baseline={([yshift=-8pt]current bounding box.north)},thick] \filldraw [fill=black] (0,0) circle (.15);
and the hole 
\tikz[baseline={([yshift=-8pt]current bounding box.north)}] \filldraw [fill=white] (0,0) circle (.15);
correspond to a negative slope edge
\begin{tikzpicture}[baseline={([yshift=-10pt]current bounding box.north)},rotate=-45,scale=.3]
\draw (0,0) rectangle ++(1,1);
\draw [very thick] (0,1) -- ++(1,0);
\end{tikzpicture}
and a positive slope edge 
\begin{tikzpicture}[baseline={([yshift=-10pt]current bounding box.north)},rotate=45,scale=.3]
\draw (0,0) rectangle ++(1,1);
\draw [very thick] (0,1) -- ++(1,0);
\end{tikzpicture}
, respectively, the derivative of the profile function $\Omega'(x)$ is given by (minus of) the density function $\rho(x)$ with a constant,
\begin{align}
    \Omega'(x) = 1 - 2 \rho(x)
    \ \implies \
    \Omega(x) = 
    \begin{cases}
    \displaystyle
    \frac{2}{\pi} \qty( x \arcsin \frac{x}{2} + \sqrt{4 - x^2}) & ( |x| < 2 ) \\
    |x| & (|x| > 2)
    \end{cases}
\end{align}

Similarly we can derive the limit shape $\Omega(x)$ for generic Schur measure from the spectral curve with the corresponding algebraic function~\eqref{eq:H_fn}.
Assuming $t_n = \tilde{t}_n$, we obtain
\begin{align}
    \qty[\sum_{n = 1}^\infty n^2 t_n \qty(y^n - y^{-n})] \frac{dy}{y} = dx
    \, .
\end{align}
Therefore, the branch point of the spectral curve $(dx = 0)$ is given by
\begin{align}
    y = 1
    \quad \xrightarrow{H(x,y) = 0} \quad
    x = \beta
    \, .
    \label{eq:sp_edge}
\end{align}
This shows that the spectral edge is in general given by $x = \beta$, and the scaling variable~\eqref{eq:sc_var} describes the fluctuation in the vicinity of the spectral edge.

\section{Discussion}

In this paper, we have discussed the universal edge scaling limit of the random partition characterized by the Schur measure.
We have in particular focused on the wave function, which constructs the kernel for the correlation function, and discussed its asymptotic behavior in the scaling limit.
We have shown that the wave function is asymptotic to the higher-order Airy function with the proper scaling variable, and thus the corresponding kernel is given similarly by the higher-analog of the Airy kernel (the $p$-Airy kernel).
As a consequence, we have then discussed that the distribution of the largest ``eigenvalue'' is described by the higher-order Tracy--Widom distribution given as the Fredholm determinant of the $p$-Airy kernel, and addressed its implication to the possible multicritical behavior in the large $N$ matrix model.
We have also discussed that the limit shape of the random partition is also concisely obtained from the semi-classical analysis of the differential/difference equation for the wave function.

Let us address several possible future directions.
In this paper, we have imposed the condition for the Schur measure parameters $\mathsf{X} = \mathsf{Y}$ to discuss the scaling limit of the kernel for simplicity.
It seems natural to relax this condition to obtain the higher-order Airy kernel for $p \in 2 \mathbb{N} + 1$.
In fact, the case with $p = 3$ is studied in the context of the gap closing regime of the random matrix theory, and the corresponding kernel is known as the Pearcey kernel~\cite{Brezin:1998zz,Brezin:1998PREb}.
In the analogy with random matrix, the current situation corresponds to the Gaussian unitary ensemble (GUE).
Then, we could also discuss the universal behavior corresponding to the Gaussian orthogonal/symplectic ensemble (GOE/GSE).
Furthermore, there is a two-parameter generalization of the Schur measure, which is called the Macdonald measure~\cite{Borodin:2014PTRF}.
It seems also interesting to study the higher-order scaling limit of these situations.
In addition, the higher-order scaling behavior discussed in this paper could be observed in experiments, e.g., the turbulent interfaces~\cite{Takeuchi:2010PRL,Takeuchi:2011SR}.
In order to realize such a higher scaling, it would be important to manipulate the merging interfaces.
It would be also interesting to further study possible implications of the asymptotic results obtained in this paper for the phase structure of the gauge theories with partition function as the unitary matrix integral~\eqref{pf}.

\subsubsection*{Acknowledgments}

This work has been supported in part by ``Investissements d'Avenir'' program, Project ISITE-BFC (No.~ANR-15-IDEX-0003), and EIPHI Graduate School (No.~ANR-17-EURE-0002).

\bibliographystyle{utphys}
\bibliography{ref}

\end{document}